\documentclass{ws-mpla}

\usepackage{psfig}

\newcommand{\los}{\mbox{{\boldmath$\ell$}}} 
\def\bB{\mbox{\boldmath$B$}}
\def\bF{\mbox{\boldmath$F$}}
\def\bDelta{\mbox{\boldmath$\Delta$}}
\def\bS{\mbox{\boldmath$S$}}
\def\bv{\mbox{\boldmath$v$}}
\def\br{\mbox{\boldmath$r$}}
\def\bg{\mbox{\boldmath$g$}}

\newcommand{\bk}{\mbox{{\boldmath$k$}}}
\newcommand{\bx}{\mbox{{\boldmath$x$}}}
\newcommand{\sbk}{\mbox{{\boldmath\scriptsize$k$}}} 
\newcommand{\sbx}{\mbox{{\boldmath\scriptsize$x$}}} 
\newcommand{\id}{\mbox{$\rm d$}} 
\newcommand{\bxi}{\mbox{{\boldmath$\xi$}}} 
\newcommand{\bnabla}{\mbox{{\boldmath$\nabla$}}} 

\begin{document}

\markboth{L.~Gizon}{Tomography of the Solar Interior}


\title{TOMOGRAPHY OF THE SOLAR INTERIOR}

\author{\footnotesize L. GIZON}

\address{Max-Planck-Institut f\"{u}r Sonnensystemforschung\\
37191 Katlenburg-Lindau, Germany\\
gizon@mps.mpg.de}

\date{Modern Physics Letters A, {\bf 21}, 1701-1715 (2006)}

\maketitle

\pub{Received 8 July 2006}{}

\begin{abstract}
Solar oscillations consist of a rich spectrum of internal acoustic waves and surface gravity waves, stochastically excited by turbulent convection. They have been monitored almost continuously over the last ten years with high-precision Doppler images of the solar surface. The purpose of helioseismology is to retrieve information about the structure and the dynamics of the solar interior from the frequencies, phases, and amplitudes of solar waves. Methods of analysis are being developed to make three-dimensional images of subsurface motions and temperature inhomogeneities in order to study convective structures and regions of magnetic activity, like sunspots.

\keywords{Sun; helioseismology; solar interior}

\end{abstract}


\section{Solar Oscillations}

The five-minute solar oscillations were discovered by Leighton, Noyes and Simon\cite{Leighton1962} and  interpreted as deep standing acoustic waves by Ulrich.\cite{Ulrich1970} Improved observations by Deubner\cite{Deubner1975} confirmed that power is concentrated along discrete ridges in wavenumber-frequency space, as predicted by Ulrich's theory. The source of excitation of solar oscillations is known to be near-surface turbulent convection.\cite{Goldreich1977,Stein2001} Standard texts on solar and stellar oscillations are listed in Refs.~\cite{Cox1980}-\cite{JCD2002}. 

The fundamental data of modern helioseismology are Doppler images of the line-of-sight component of velocity, $\Phi(\bx,t)$, where $t$ is time and $\bx$ is a position vector on the Sun's surface. The two main datasets are provided by the Global Oscillation Network Group (GONG, Ref.~\cite{Leibacher1999}) and by the Michelson Doppler Imager (MDI, Ref.~\cite{Scherrer1995}) aboard the ESA/NASA SOHO spacecraft in a halo orbit around the L1 Sun-Earth Lagrange point since 1996. The MDI Dopplergrams are obtained by combining four filtergrams in the wings and core of the Ni 6788~{\AA} absorption line, formed just above the photosphere; images are recorded on a 1024$\times$1024 pixel CCD camera at the cadence of one per minute. 

\begin{figure}[t]
\centerline{\psfig{file=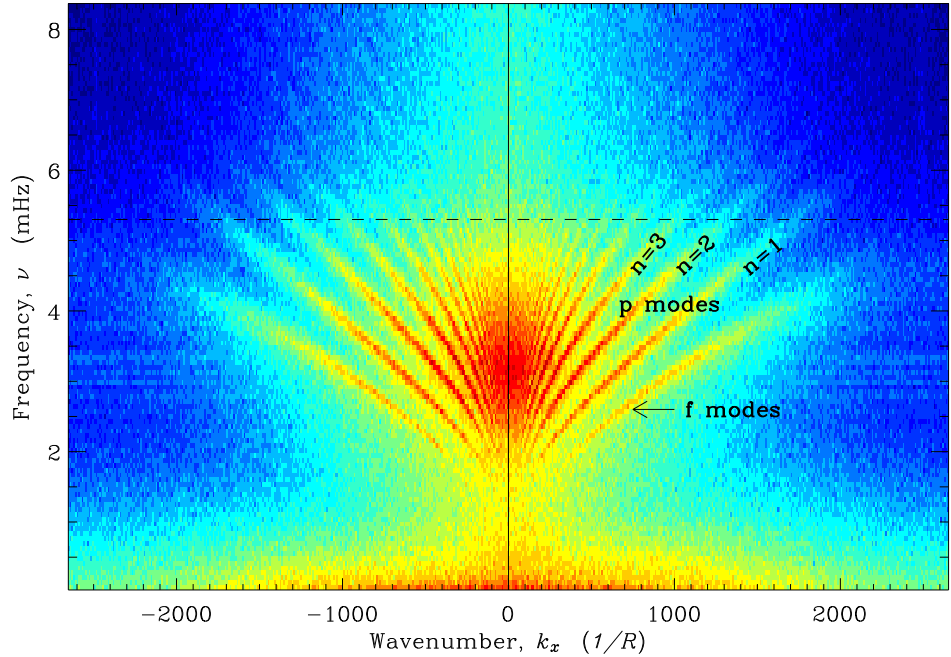,width=5.5in,clip=}}
\caption{Cut at $k_y=0$ through a three-dimensional power spectrum of solar oscillations, $P(\bk,\omega)$, obtained by the MDI-SOHO instrument in its high-resolution mode of operation. This power spectrum is an average over 8 individual power spectra (the total period of observation is 8 times $T=4$~hr). The wavenumber is given in units of the inverse solar radius. The dashed lines indicates the acoustic cutoff frequency at $5.3$~mHz, above which acoustic waves are not trapped inside the Sun.}
\label{fig.power}
\end{figure}

The data are best studied in wavenumber-frequency space. Under the assumption that the Sun's surface is locally plane, we can extract the harmonic components of the signal by application of a 3D Fourier transform:
\begin{equation}
\tilde{\Phi}(\bk,\omega) = \int_{A} \id^2\bx \int_0^T  \id t \; \Phi(\bx,t) e^{i\sbk\cdot\sbx - i \omega t} ,
\end{equation}
where $A$ is a local area on the Sun, $\bk=(k_x,k_y)$ is the horizontal wavevector, and $\omega=2\pi \nu$ is the angular frequency. We adopt the convention that the $x$ coordinate points prograde, toward the west limb, while $y$ points toward the north pole. Figure~\ref{fig.power} shows a cut at $k_y=0$ through the power spectrum $P(\bk,\omega)=|\tilde{\Phi}|^2$ obtained from a time series of MDI Doppler images. Images were previously shifted in space to remove the main component of solar rotation. 

Power is distributed along distinct ridges labeled by an integer $n$, known as the radial order, which corresponds to the number of nodes of the wave eigenfunctions counted in the radial direction (from the surface to the center of the Sun). The lowest frequency ridge with $n = 0$ corresponds to the fundamental (f) modes of solar oscillations. The f modes are identified as surface gravity waves, with nearly the dispersion relation for deep water waves, $\omega^2  = gk$, where $g = 274$~m\,s$^{-2}$ is the gravitational acceleration at the Sun's surface and $k$ is the horizontal wavenumber. The f modes propagate horizontally. All other ridges, denoted by p$_n$, correspond to acoustic modes, or p modes. The restoring force for p modes is pressure. The ridge immediately above the f mode ridge is p$_1$, the next one p$_2$, and so forth.   We may denote by $\omega = \omega_n(\bk)$ the dispersion relation associated with the $n$th ridge of power. Modes with smaller values of $k$ and larger $n$ penetrate deeper inside the Sun. An essential physical property of acoustic waves is that modes with similar values of the horizontal phase speed $\omega_n(k) / k$ propagate to similar depths inside the Sun. For frequencies above the acoustic cutoff frequency ($5.3$~mHz, dashed line), acoustic waves are not trapped inside the Sun but propagate freely outward.   We note that for $k<150/R_\odot$, where $R_\odot=696$~Mm is the solar radius, it is not correct to assume a plane parallel geometry: spherical harmonics transforms should be implemented instead of spatial Fourier transforms, as is done in global mode helioseismology.\cite{JCD2002}

The theory of solar oscillations has been discussed at length elsewhere.  In local helioseismology, it is important to consider wave propagation through generic solar models, including models with local inhomogeneities. In this paragraph and for the sake of simplicity only, we assume that no steady background flow  is present. Solar oscillations are described by the displacement vector $\bxi(\br,t)$ of a fluid parcel that would have been at location $\br$ and time $t$ had there been no wave motion. All physical quantities (e.g. density) are expanded into a steady background value (e.g. $\rho$)  and a wave perturbation (denoted with a prime, e.g. $\rho'$). In the case of small-amplitude oscillations, which is a good approximation, only first order perturbations are retained. The continuity equation reads
\begin{equation}
  \rho' + \bnabla \cdot ( \rho \bxi) = 0 .
\label{eq.continuity}
\end{equation}
Ignoring the perturbation to the gravitational potential (Cowling approximation), the momentum equation is
\begin{equation}
\rho \partial_t^2 \bxi = - \bnabla p' + \rho' \bg ,
\label{eq.momentum}
\end{equation} 
where $p'$ is the pressure perturbation and $\bg = (\bnabla p) / \rho$ is the gravitational acceleration. We note that if the background model includes a magnetic field $\bB$ then the perturbation to the Lorentz force, $\bF[\bxi] = (\bnabla\times\bB)\times\bB'/4\pi + (\bnabla\times\bB')\times\bB/4\pi$, must be added to the RHS of Eq.~(\ref{eq.momentum}). For the energy equation, we assume adiabatic wave motion (no heat losses):
\begin{equation}
\frac{ \delta p}{p} = \Gamma_1 \frac{ \delta\rho }{\rho}   = - \Gamma_1 \bnabla \cdot \bxi ,
\label{eq.energy}
\end{equation}
where $\Gamma_1$ is the first adiabatic exponent and $\delta p = p' + \bxi\cdot\bnabla p$ and  $\delta \rho = \rho' + \bxi\cdot\bnabla \rho$ are the Lagrangian perturbations in pressure and density respectively. Combining the above equations, we obtain a single differential equation for $\bxi$:
\begin{equation}
- \rho \partial_t^2 \bxi + \bnabla( \Gamma_1 p \bnabla\cdot\bxi + \bxi\cdot\bnabla p) - (\bnabla\cdot\bxi + \bxi\cdot\bnabla) \bnabla p  +   \rho (\bxi\cdot\bnabla) \bg  = 0.
\label{eq.comb}
\end{equation}
 Had we considered a medium with a background flow, $\bv$, then the time derivative $\partial_t$ in the above equation would have been replaced by $\partial_t + \bv\cdot\bnabla$, as explained in Ref.~\cite{LyndenBell1967}. We note that, in the upper layers of the Sun, the term $\rho (\bxi\cdot\bnabla) \bg$ can usually be dropped because variations in $\bg$ are slow there. 

The point of the previous paragraph was to show that it is feasible to write down a differential operator, ${\cal L}$, that controls the evolution of the fluid displacement associated with free, small-amplitude oscillations of the Sun. Our model, however, is incomplete as it ignores the excitation and damping of oscillations by turbulent convection, a difficult problem. It is customary to add a source term, $\bS$, to the RHS of Eq.~(\ref{eq.comb}), 
\begin{equation}
{\cal L} [ \bxi ]  = \bS(\br,t) ,
\label{eq.main}
\end{equation}
and to tune $\bS$ empirically to obtain an approximate match between the model and observed power spectra.\cite{Birch2004} At the same time, a simplified damping term is also added to ${\cal L}$ to reproduce the linewidths of the modes in the power spectrum.\cite{Birch2004} The function $\bS$ should be understood as the realization of a random process. 

By definition, the observable is the line-of-sight component of velocity at the solar surface:  $\Phi(\bx,t)  = {\rm psf}\otimes( \los \cdot \partial_t \bxi)$, where $\los$ is a unit vector that points toward the observer and ${\rm psf}\otimes$ denotes the convolution of the signal by the point spread function of the telescope. 

\section{Methods}

Local helioseismology is a set of tools to make three dimensional images of the solar interior. The most popular methods of analysis are time-distance helioseismology, ring-diagram analysis, Fourier-Hankel decomposition, seismic holography, and direct modeling (see Ref.~\cite{Gizon2005} for a general review). Here I shall only explain the basic principle of time-distance and ring-diagram analyses.

Time-distance helioseismology\cite{Duvall1993} consists of measuring the travel times of wave packets propagating  through the solar interior  between any two points on the solar surface. The travel times between locations $\bx-\bDelta/2$ and $\bx+\bDelta/2$ are extracted from the cross-covariance function 
\begin{equation}
C(t) = \int_0^T \Phi(\bx-\bDelta/2,t') \Phi(\bx+\bDelta/2,t'+t) \id t' .
\end{equation}
The cross-covariance function is a phase-coherent average of the random oscillations: it can be seen as a solar seismogram.  The travel time $\tau(\bx, \bDelta)$ for waves moving from $\bx-\bDelta/2$ to $\bx+\bDelta/2$  is measured by fitting $C(t>0)$ with a Gaussian wavelet. A travel time anomaly contains the seismic signature of buried inhomogeneities within the proximity of the ray path that connects the two surface locations. Flows in the Sun break the symmetry between $\tau(\bx,\bDelta)$ and $\tau(\bx,-\bDelta)$, while sound speed perturbations affect the mean travel time. An inverse problem must be solved to infer the 3D structure and dynamics of the solar interior.

Ring-diagram analysis\cite{Hill1988} consists of studying local power spectra  of solar oscillations, $P(\bk,\omega; \bx)$, computed over small patches of the solar surface around a location $\bx$ (unlike the power spectrum of Figure~\ref{fig.power}, which was computed over a large fraction of the solar disk). The frequencies of the modes of solar oscillations are carefully measured from any given local power spectrum. The small differences between the measured mode frequencies and the theoretical frequencies from a reference solar model provides information about the solar interior in the vicinity of $\bx$. Ring-diagram analysis has been particularly successful at mapping horizontal flows in the upper convection zone. The effect of a slowly-varying horizontal flow $\bv$ is to Doppler shift the power spectrum by an amount $\bk\cdot\bv$:
\begin{equation}
P(\bk,\omega; \bx) = P_0(\bk,\omega-\bk\cdot\bv) ,
\end{equation}
where $P_0$ is the reference power spectrum with $\bv=0$. In practice, for a general flow $\bv(\br)$ that varies with depth, the Doppler shift will depend on the radial order. Once again, an inverse problem must be solved to infer the variation of solar properties with position in the solar interior.

In order to do linear inversions of helioseismic data, it is necessary to first solve the linear forward problem.  The linear forward problem is to compute the first-order effect (prefix $\delta$) of small perturbations to a reference solar model (subscript $0$). The zero-order reference solar model is usually invariant by horizontal translation, has zero mean flow, and no magnetic field.     In general, the linear forward problem can be written as
\begin{equation}
\delta d (\bx) = \sum_\alpha \int_\odot\id^3\br\; K_\alpha(\br; \bx) \delta q_\alpha(\br) \, .
\end{equation}
In this equation, $\delta d$ denotes the value of a particular seismic parameter (travel time, ring fit parameter, etc...)  minus the value of this parameter in the reference solar model. The sum over the index $\alpha$ is a sum over all possible types of physical quantities $q_\alpha$ (sound speed, density, flows, etc...). The kernel function  $K_\alpha(\br; \bx)$ gives the sensitivity of $\delta d(\bx)$ to the perturbation $\delta q_\alpha(\br)$ at position $\br$ in the solar interior. So far, essentially all studies assume that the perturbations to the solar model, $\delta q_\alpha$, are time-independent over the time duration during which the observations are made.

The main step involved in the derivation of the kernels $K_\alpha$ is the calculation of the first order perturbation to the wave displacement, $\delta\bxi$. The change in the solar model, through the $\delta q_\alpha$, are included in the perturbed wave operator, $\delta{\cal L}$. The $\delta q_\alpha$ may also imply a change in the source function, $\delta \bS$. For example, a magnetic field or a flow may affect the small-scale convection that drives the solar oscillations. The calculation of $\delta \bxi$, according the first-order Born approximation,
\begin{equation}
{\cal L}_0 [ \delta \bxi ] = - \delta{\cal L} [ \bxi_0 ] + \delta \bS ,
\label{eq.Born}
\end{equation}
 is developed in full detail in Refs.~\cite{Gizon2002}-\cite{Birch2004}. Equation~(\ref{eq.Born}) shows that the Born approximation is an equivalent-source description of wave interaction. An equally good alternative to the Born approximation is the Rytov approximation.\cite{Jensen2003}

\begin{figure}[t]
\centerline{\psfig{file=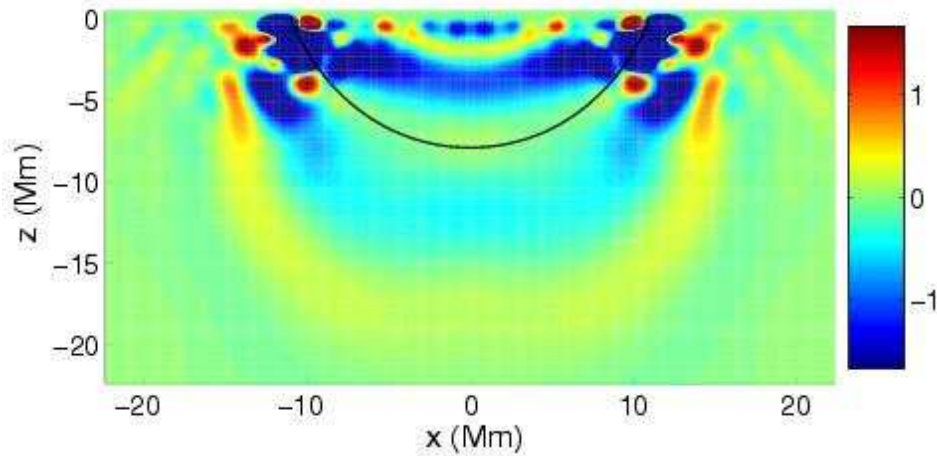,width=4.95in, clip=}}
\caption{Cut through a three dimensional kernel giving the sensitivity of travel times to sound speed perturbations ($\delta q_\alpha = \delta c^2 / c^3$). The scattering of acoustic waves by sound speed inhomogeneities was calculated under the first-order Born approximation. The ray path (infinite frequency limit) is given by the black line. The distance between the two points at the surface is $\Delta=20$~Mm and the units of the kernel are $10^{-2}$~Mm.  }
\label{fig.kernel}
\end{figure}

Figure~\ref{fig.kernel} shows a kernel function obtained by Birch, Kosovichev and Duvall\cite{Birch2004} for the sensitivity of travel-time measurements ($\delta d = \delta\tau$) to a local perturbation $\delta q_\alpha = \delta c^2 / c^3$, where $c=(\Gamma_1 p /\rho)^{1/2}$ is the sound speed. The kernel values are close to zero along the ray path (infinite frequency limit), which led geophysics to refer to such kernels as banana-doughnut kernels.\cite{Dahlen2000}

The sensitivity of travel times extends much beyond the geometrical ray path that connects the two points at the surface. The central lobe of a sensitivity kernel is called the first Fresnel zone: a scatterer placed on the boundary of the first Fresnel zone causes a phase difference between the direct and scattered waves of $\pi$. In the case of a uniform medium, the width of the Fresnel zone is approximately given by $L_F \sim (\lambda \Delta)^{1/2}$, where $\lambda$ is the dominant wavelength of the solar oscillations. To illustrate this point, we may take as a reference for $\lambda$ the wavelength of an f mode near $3$~mHz:   
\begin{equation}
L_F \sim 10~{\rm Mm} \; \left( \frac{\lambda}{5~{\rm Mm}}\right) ^{1/2}\left( \frac{\Delta}{20~{\rm Mm}}\right) ^{1/2}  .
\end{equation}
The width of the first Fresnel zone does exceed the size of some of the solar convective structures that leave a signature in the travel-time measurements. To be more precise, finite-wavelength effects become very important when the wavelength of the underlying perturbations in the Sun is less than $2 L_F$.  We note that finite-wavelength tomography has received a lot attention in geophysics in the last few years, in particular to resolve plumes in the Earth's mantle.\cite{Montelli2004} One of the differences between helio- and geo-seismology is that in the Sun waves are excited stochastically over the whole solar surface, leading to more complex sensitivity kernels. 

The linear inverse problem of local helioseismology, which consists of inverting for the $\delta q_\alpha$ from a set of measurements $\delta d$, is relatively well understood. Two standard methods are used: regularized least squares (RLS, Ref.~\cite{Hansen1998}) and optimally localized averaging (OLA, Ref.~\cite{Backus1968}).  The multi-channel deconvolution algorithm significantly speeds up RLS inversions.\cite{Jensen2001} In general, inversions must take into account the fact that the data errors are correlated.\cite{Jensen2003,Gizon2004,Couvidat2005}

\section{Convection}

Motions on the solar surface display two specific scales: granulation and supergranulation. Granules, with a typical scale of $1.5$~Mm, are well understood as a convective phenomenon thanks to realistic numerical simulations. Supergranules, however, have remained puzzling since their discovery by Leighton, Noyes and Simon:\cite{Leighton1962} there is no accepted theory that explains why convection should favor a 30~Mm scale. Supergranular cells expel the magnetic flux from the regions of fluid motion and concentrate it into ropes at the cell boundaries to form the quiet Sun magnetic network.

\begin{figure}[th]
\centerline{\psfig{file=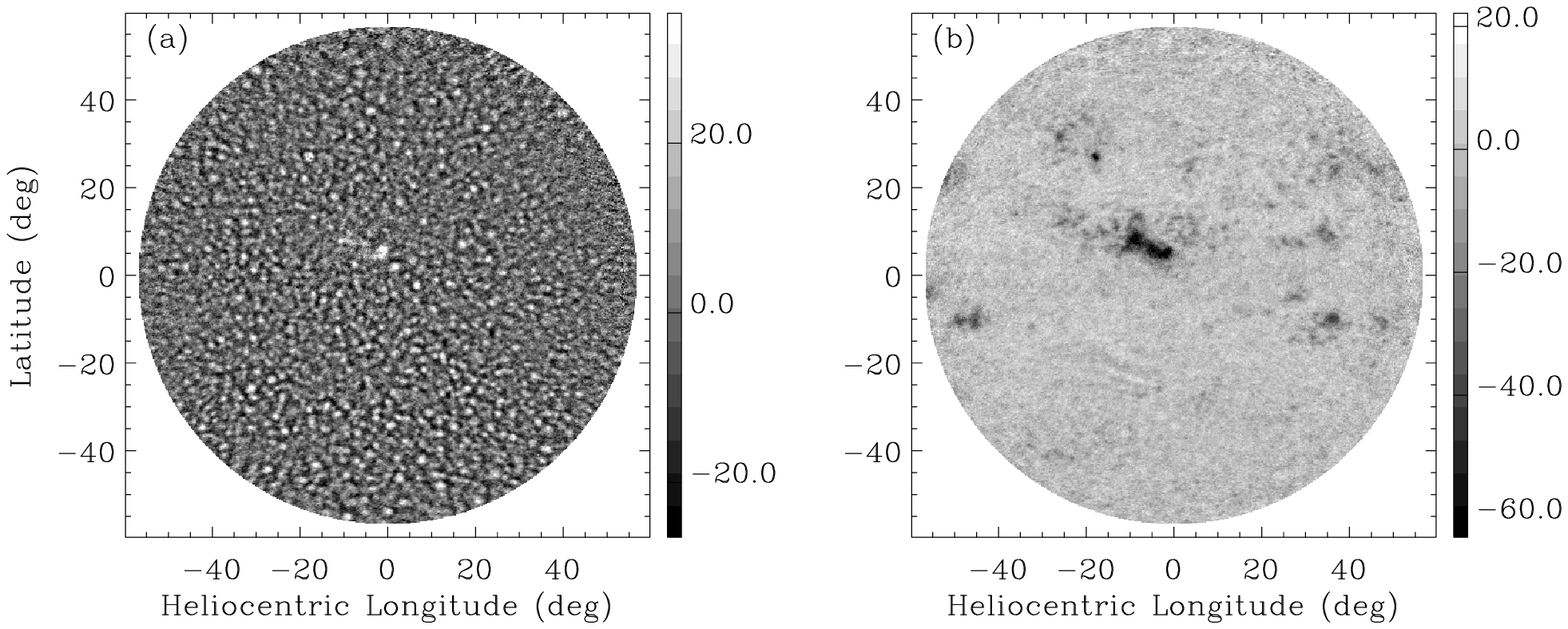,width=5in,clip=}}
\caption{Map of the horizontal divergence of the flows in a 1-Mm deep layer below the photosphere. Center-to-annulus f-mode travel-time differences are given in units of second (observation duration $T=12$~hr, annulus radius $\Delta=15$~Mm, MDI full-disk data). A positive signal corresponds to a positive divergence.}
\label{fig.oi}
\end{figure}

The simplest way to detect supergranulation with time-distance helioseismology is to measure, with a cross-correlation technique, the time it takes for solar waves to propagate between any given point on the solar surface and a concentric annulus around that point. The difference in travel times between inward and outward propagating waves is a measure of the local divergence of the flow.\cite{Duvall1997} In the case of surface-gravity waves (f modes), the travel-time difference is directly sensitive to the horizontal divergence of the flow velocity in the top 2~Mm below the surface, without inversion.\cite{Duvall2000} Figure~\ref{fig.oi} shows the divergence signal (inward minus outward travel times) obtained by analyzing a 12-hr long time series of MDI full-disk Doppler velocity images tracked at the Carrington rate to remove the main component of solar rotation. A white, or positive signal, corresponds to a horizontal outflow. From the size of the features present and their locations with respect to the magnetic network, supergranulation is identified as the main contribution to the signal.  Compared to the Doppler line-of-sight velocity, the main advantage of the helioseismic divergence signal is that it is essentially free of systematic variations across the field of view, which is an important condition to study the evolution of the pattern. Divergence maps can also be constructed with a technique known as seismic holography.\cite{Braun2003}

The evolution of the pattern of convection has been studied using long sequences (two to three months each year) of divergence maps, like the one shown in Figure~\ref{fig.oi}. Each map covers a region of size $120^\circ\times120^\circ$ centered on the solar disk. This means that a feature co-rotating with the Sun can be observed continuously for about $9$~days. Such long time series enabled Gizon, Duvall and Schou\cite{Gizon2003} to resolve unexpected oscillations in the 3D Fourier spectrum of the divergence signal (with frequencies in the range $1$-$2$~$\mu$Hz) and excess power in the prograde direction. These wave-like properties have not been  explained yet, although an interesting analysis by Green and Kosovichev\cite{Green2006} suggests that convection in a vertical shear flow could take the form of traveling waves. From the autocorrelation of the divergence signal, the lifetime of supergranulation is measured to be about $2$~days. 

In addition to the horizontal divergence, ${\rm div} = \partial_x v_x +  \partial_y v_y$, it is also possible to measure separately the two horizontal components of the vector flow (by splitting the annulus into quadrants), from which an estimate of the vertical vorticity, ${\rm curl}= \partial_x v_y - \partial_y v_x$, can be derived. After removing the large-scale vorticity\cite{Zhao2004} due to differential rotation and meridional circulation, a small correlation between div and curl is detected,\cite{Duvall2000,Gizon2003b} which is found to be  proportional to $\Omega(\lambda)\sin\lambda$, where $\Omega(\lambda)$ is the angular velocity of the Sun at latitude $\lambda$. This latitudinal variation of the correlation is precisely what is expected of the effect of the Coriolis force on convection. In the northern hemisphere, supergranular cells rotate preferentially clockwise where the horizontal divergence is positive, while they rotate preferentially counterclockwise in the convergent flows near the sinks. The sense of circulation is reversed in the southern hemisphere. The coupling between convection and rotation is weak (the correlation coefficient between div and curl is less than $2\%$), in relatively good agreement with numerical simulations.\cite{Hathaway1982,Egorov2004}

In addition, both time-distance helioseismology and ring diagram analysis have revealed the existence of long-lived flow patterns on scales that are much larger than supergranulation,\cite{Gizon2001,Haber2001} which may be related to the largest scales of deep convection.\cite{Toomre2002}

\section{Sunspots and Active Regions}
\label{sec.ar}

Thomas, Cram and Nye\cite{Thomas1982} first suggested that solar oscillations could be used to probe the internal structure of sunspots. Since then, observations have shown that the frequencies, phases, and amplitudes of solar oscillations are perturbed in regions of enhanced magnetic activity.\cite{Bogdan2000} Sunspots are known to absorb incident p-mode energy\cite{Braun1987}, introduce phase shifts between the incident and transmitted waves\cite{Braun1992,Duvall1993}, and cause mode mixing.\cite{Braun1995} All these observations inform us about the conditions inside and in the immediate vicinity of sunspots: temperature and density anomalies, fluid motions, and the magnetic field.\cite{Duvall1996,Kosovichev1996,Cally2003} High-frequency acoustic waves can also  be used to map the topography of the magnetic fields above sunspots, in the chrosmosphere.\cite{Finsterle2004}

\begin{figure}[th]
\centerline{\psfig{file=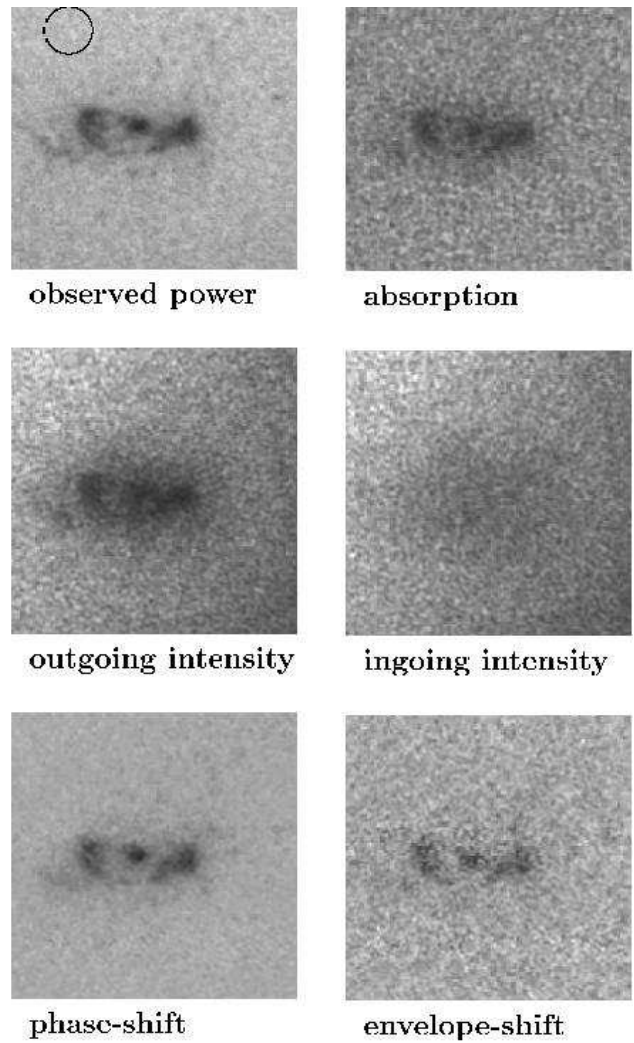,width=3.9in, clip=}}
\vspace*{8pt}
\vspace{-0.5cm}
\caption{Acoustic imaging of active region NOAA~7978 using a $33$~hr MDI time series of Doppler maps. The panels shows reduced p-mode power, enhanced p-mode absorption (deficit of outgoing intensity), and decreased phase and envelope travel times at the location of the active region. The data were filtered with a Gaussian filter centered around frequency $3.5$~mHz. The acoustic power is reduced by a factor of four in the central sunspot. The size of the aperture used is $2$-$6^\circ$ (a $2^\circ$ circle is shown in the upper left map). The dimension of each map is $24.0^\circ$ in longitude and $24.7^\circ$ in latitude.}
\label{fig.dean}
\end{figure}

Figure~\ref{fig.dean} shows results obtained by Chou\cite{Chou2000} with acoustic imaging, a technique used to reconstruct the seismic signal at any target point on the solar surface. The reconstructed signal contains information about the amplitude and the phase of wave packets converging toward the target point or diverging from that point. Active regions introduce local changes easily detectable in the maps of the reconstructed outgoing intensity.  Waves are absorbed by active regions as evidenced by the deficit of outgoing acoustic intensity.  In this example, the active region also introduces phase-time shifts as big as $1.5$~min, i.e. a significant fraction of the wave oscillation period.   

The forward and inverse problems of sunspot seismology present many challenging issues. The main difficulties are nonlinear aspects of wave propagation, radiative transfer in magnetized plasmas, and the relationship between velocity measurements in sunspots and real fluid motions. The complexity of the problem is evident from numerical studies of the propagation of p and f modes through model sunspots and their coupling to magnetic waves\cite{Cally1997,Cally2000}. It should be noted that magnetic perturbations are not small near the solar surface:\cite{Lindsey2004,Cally2005} the first Born approximation cannot be applied there. Only deeper inside the Sun, can magnetic effects be treated as small perturbations.\cite{Gizon2006b} 

\begin{figure}[th]
\centerline{\psfig{file=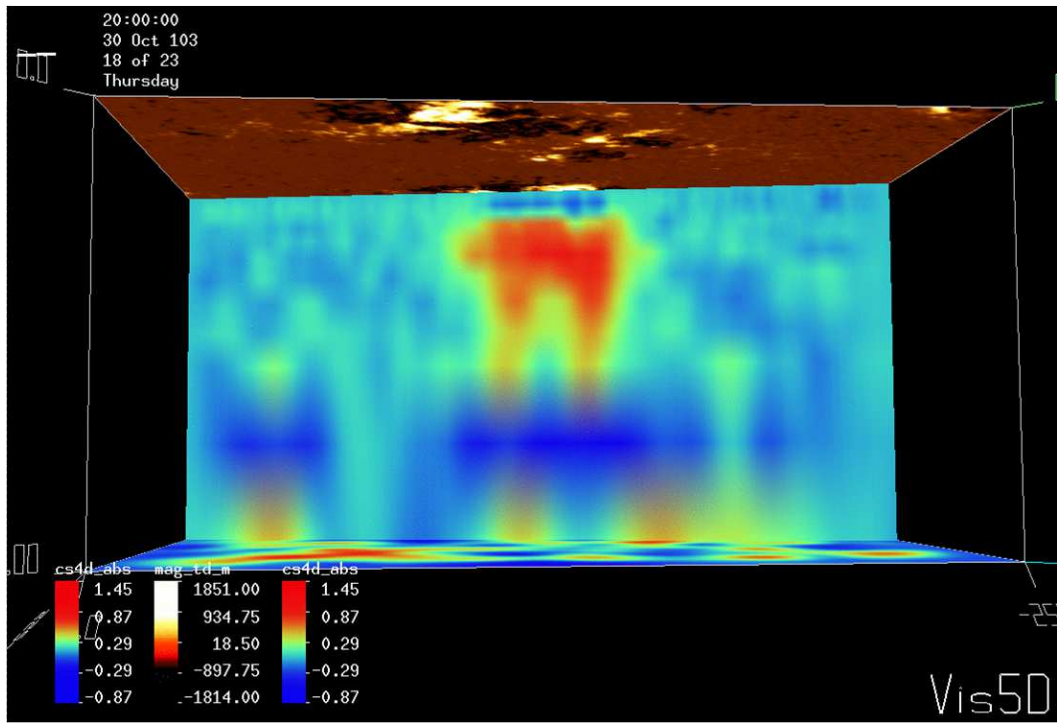,width=5.5in, clip=}}
\caption{Vertical cut through the 3D sound-speed variations associated with active region NOAA 10488. The depth of the box is 48~Mm, the horizontal size is 540~Mm. The sound-speed scale ranges from $-1$~km$\,$s$^{-1}$ (blue) to $1.5$~km$\,$s$^{-1}$ (red). }
\label{fig.koso}
\end{figure}

A number of simplifications have been made to interpret the observations and obtain approximate answers.  The first assumption which is commonly used is that structural perturbations inside sunspots are small (not quite correct, as mentioned just above). In addition, it is often assumed that the travel-time shifts, scattering phase shifts, or local mode frequency shifts are caused by sound speed perturbations (indirect effect of the magnetic field). Under these simplifying assumptions, a linear inverse problem can be solved to infer subsurface sound-speed perturbations.\cite{Kosovichev1996}  One of the latest inversions by Kosovichev\cite{Kosovichev2004} using high-resolution MDI-SOHO data, is shown in Figure~\ref{fig.koso}. It is found that the sound speed is lower just below the sunspots (probably indicating a lower temperature), while the sound speed is higher at depths greater than about 4~Mm. Deeper than about $15$~Mm, the seismic signature of sunspots fades away. These findings indicate that sunspots are rather shallow and that the emerged magnetic flux is somehow disconnected from its root in the deep convection zone.\cite{Schuessler2005} Figure~\ref{fig.koso}, a vertical cut through a pair of sunspots with opposite polarities, reveals a very interesting loop-like structure, which may tell us something about the underlying magnetic topology.   

Efforts are under way to infer other quantities than the sound speed below sunspots. Basu, Antia and Bogart\cite{Basu2004} performed a depth inversion of local frequency shifts simultaneously for the sound speed and the adiabatic exponent. A major goal is to directly image the magnetic field. This has not been accomplished yet as it is not easy to separate unambiguously magnetic perturbations from other perturbations, such as temperature or density anomalies. Piercing through the surface will require that we know how to correct for the strong phase shifts introduced by the surface density drop (Wilson depression) and the surface magnetic field.\cite{Lindsey2005,Schunker2005,Zhao2006}  Ultimately, non-linear inversions of seismic measurements may be necessary. A first attempt is due to  Crouch et al.\cite{Crouch2005} using a genetic algorithm to adjust the parameters that control the radial structure of sunspot models. A major improvement in our understanding of wave propagation through strongly magnetized regions is expected from realistic numerical simulations.

\section{Thin Magnetic Tubes}

Outside active regions, in the quiet Sun, the photosphere magnetic fields appear to be clumped into intense flux tubes with typical field strength of order $1$~kG and diameters of about $100$~km. It is well known that flux tubes support various modes of oscillation: the Alfv\'en modes (twisting motion), the sausage modes (propagating change in the cross-section of the tube), and the kink modes (bending of the tube).   The interaction of acoustic waves with thin flux tubes, i.e. tubes with diameters much less than the wavelengths, has been studied extensively. In particular, it is believed\cite{Bogdan1996,Tirry2000} that scattering of acoustic waves by flux tubes contributes to the observed\cite{Rajaguru2001,Komm2002} damping rates and frequency shifts. Local helioseismology has the potential to provide observational tests for some of these ideas. A promising approach, pioneered by Duvall, Birch and Gizon,\cite{Duvall2006} is to study small magnetic features averaged over 4~hr time intervals. Compared to sunspots, magnetic network elements have the advantage of being relatively simple magnetic configurations. In addition, they are so many that a statistical treatment is possible. It was observed that the decrease in f-mode travel time caused by an average magnetic feature is $2$~s$\,$kG$^{-1}$ at most. It is hoped that such data will help us understand wave scattering by slender vertical flux tubes.

\section{Mass Flows and Magnetism}

A distinction should be made between flows in the immediate vicinity of sunspots, and small-amplitude, extended flows around large active regions. Duvall et al.\cite{Duvall1996} discovered travel-time shifts in sunspots that are consistent with strong, 1~km$\,$s$^{-1}$ downflows, several~Mm below the surface. Near the surface, a 500~m$\,$s$^{-1}$ horizontal outflow from the center of sunspots, known as the moat flow, is observed.\cite{Lindsey1996,Gizon2000,Braun2003}  Zhao, Kosovichev and Duvall\cite{Zhao2001} discovered mass motions across a sunspot at a depth of about $10$~Mm, as well as a collar flow, which may stabilize the sunspot.\cite{Botha2006}

On a much larger spatial scale (tens of heliographic degrees), long time averages of surface flow maps reveal weak, $50$~m$\,$s$^{-1}$ horizontal flows that converge toward large complexes of magnetic activity.\cite{Gizon2001,Haber2001} Such a surface inflow may have its origin in the decrease of gas pressure in magnetized regions.\cite{Spruit2003} Deeper inside the Sun ($10$~Mm below the surface), inversions point to the existence of $50$~m$\,$s$^{-1}$ horizontal flows diverging away from active regions,\cite{Haber2004,Zhao2004b} suggesting the existence of an extended toroidal cell.\cite{Gizon2004c}  Using a mass conservation constraint, Komm et al.\cite{Komm2004} derived the vertical component of velocity and the kinetic helicity density, which take the largest values in active regions.

Local helioseismology has also been particularly successful at measuring global-scale motions in the Sun: rotation and meridional circulation (See Ref.~\cite{Gizon2004c} and references therein). The localized, organized motions around active regions that have just been described introduce an 11-yr solar-cycle variation in the longitudinal average of the meridional circulation, which could affect the large-scale transport of the surface magnetic flux. 

\section{Farside Imaging}
\begin{figure}[t]
\centerline{\psfig{file=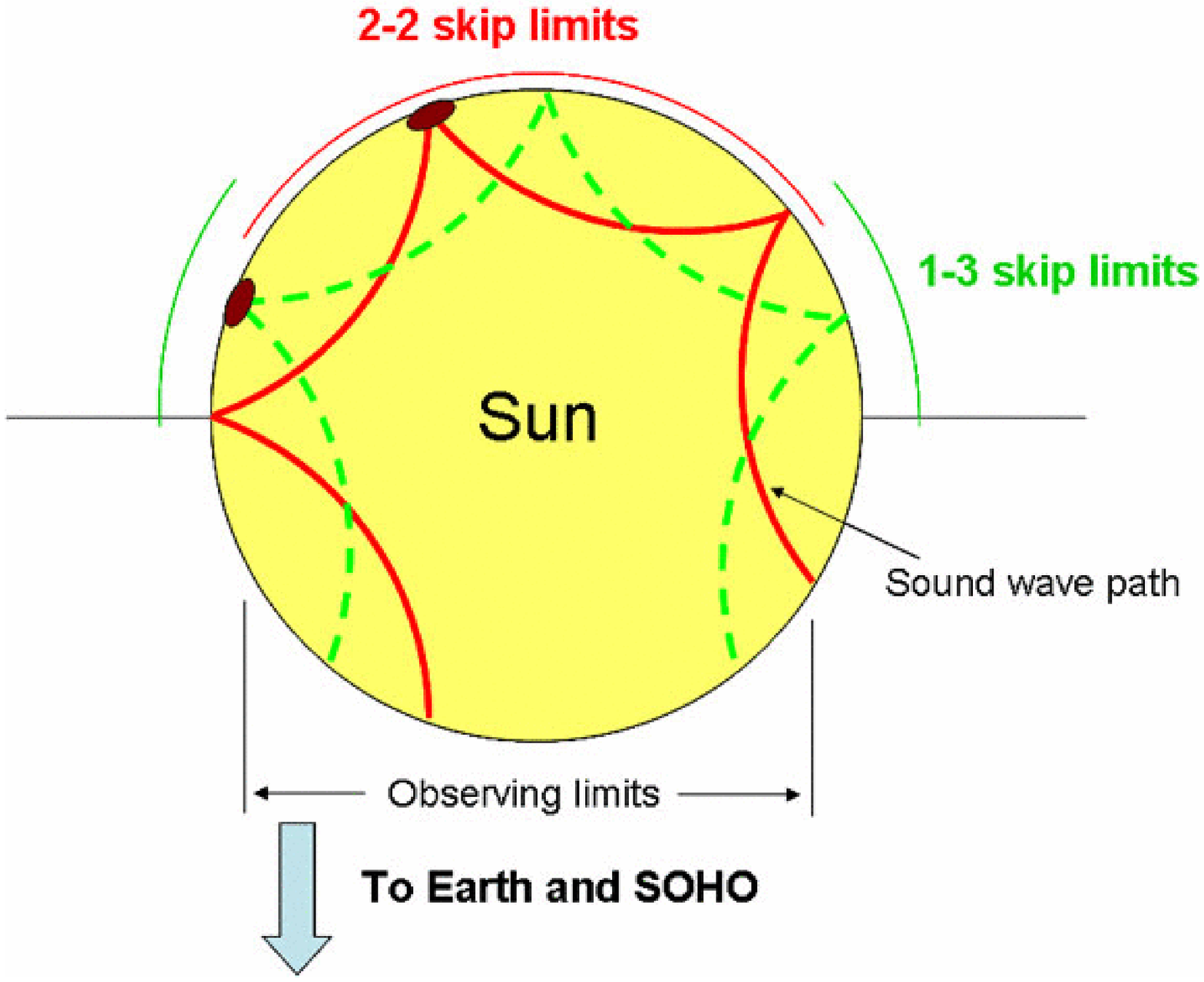,width=3.in, clip=}}
\centerline{\psfig{file=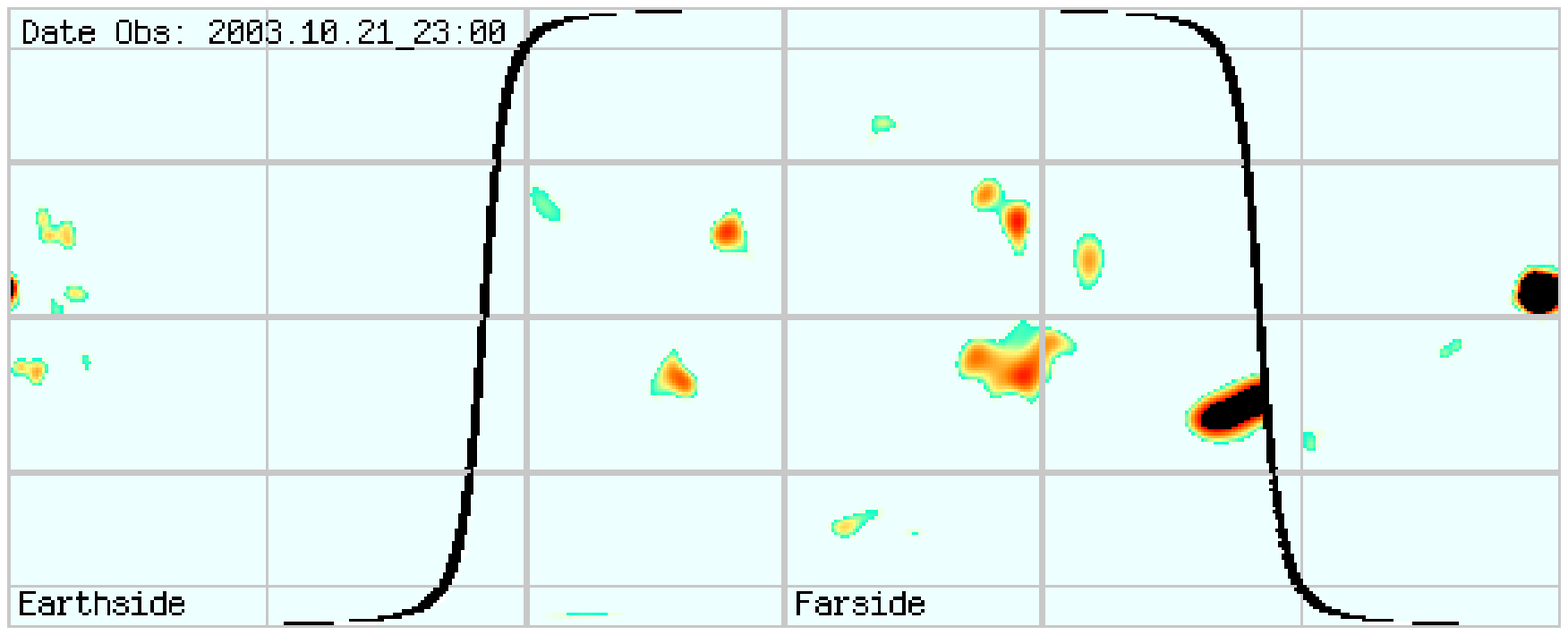,width=4.in, clip=}}
\vspace{-4.5cm}
\centerline{\psfig{file=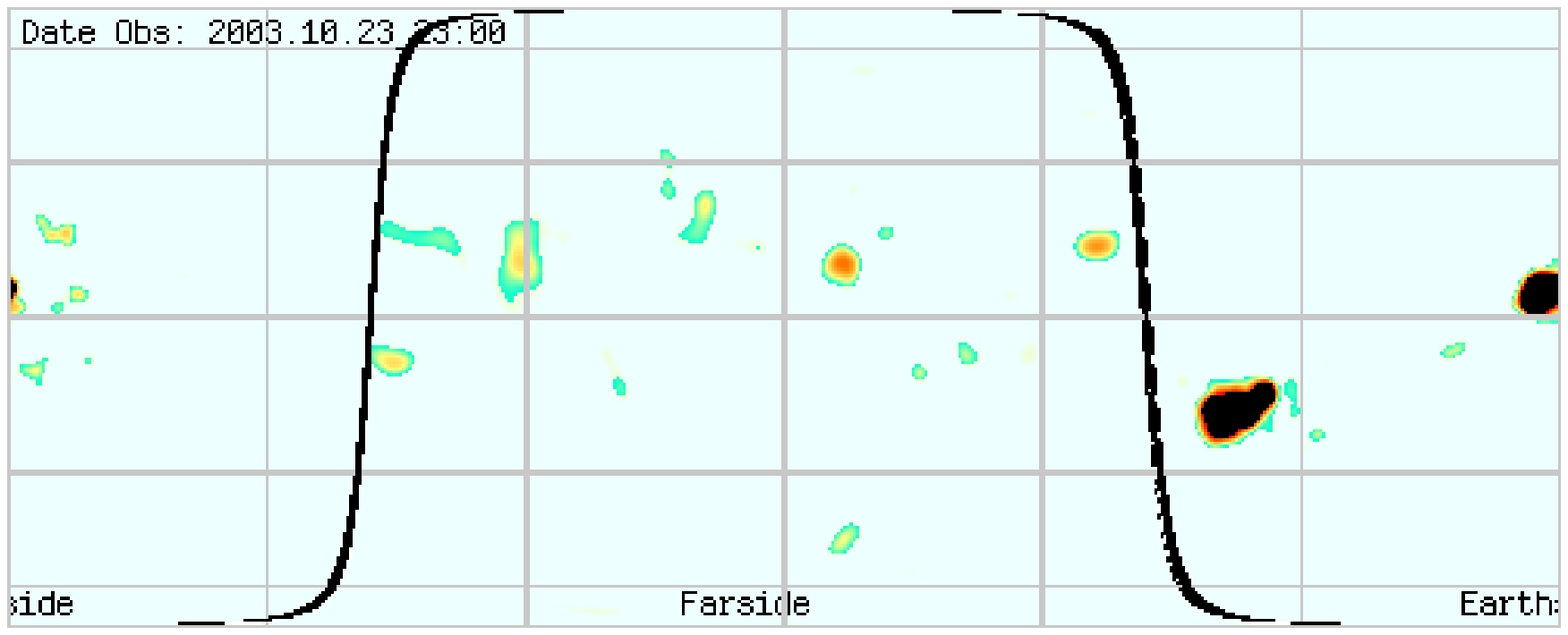,width=4.in, clip=}}
\vspace{-3.5cm}
\caption{Top: Multi-bounce ray paths connecting the front side (Earth side)  and the farside of the Sun. Acoustic waves that bounce off an active region situated on the farside suffer a measurable phase shift.  Maps of the full farside are obtained by combining the 2-2 (red), 1-3 (green), and 3-1 skip geometries. Middle and bottom panels: Farside imaging of the Sun on 22 October 2003 (middle panel) and 24 October 2003 (bottom panel). A large active region can be seen on October 22 on the farside just before it rotates onto the front side of the Sun on October 24. The black curve is the boundary between the far and front sides.}
\label{fig.farside}
\end{figure}

Lindsey and Braun\cite{Lindsey2000} demonstrated that acoustic waves observed on the front side (Earth side) of the Sun can be used to image large active regions on the far side of the Sun. Farside helioseismic imaging, a special case of phase-sensitive holography, uses long-wavelength internal acoustic waves that leave the front side, bounce once off the surface, then bounce from a target location (on the farside), and bounce again on the way returning to the front side where we can observe them again.  This 2+2 skip geometry (Figure~\ref{fig.farside}) is designed to map regions not too distant from the antipode of the center of the visible disk. Active regions on the farside introduce travel-time shifts of the order of ten seconds with respect to the quiet Sun. For focus positions closer to the limb, Braun and Lindsey\cite{Braun2001} proposed to use a 1+3 skip geometry. By combining the 2+2 and 1+3 skip geometries, Oslund and Scherrer (2006) produced maps of the entire farside (see Figure~\ref{fig.farside}). Obviously, farside imaging has important implications  for space weather predictions.

\section{Conclusion}
This short overview of the methods and results of helioseismic tomography will hopefully convince the reader of the richness of this exciting field of research. Local helioseismology has revealed unexpected aspects of the  dynamics of supergranulation, wave-speed anomalies and complex flow patterns below sunspots and active regions, and magnetic activity on the farside of the Sun. As mentioned throughout this text, many challenging issues will have to be addressed before these discoveries can be fully be trusted. Future progress is likely to come from improved models of the interaction of seismic waves with subsurface inhomogeneities and validation through realistic numerical simulations. On the observational side, the next technological steps are NASA's Solar Dynamics Observatory to be launched in 2008 (improved spatial resolution) and the Solar Orbiter mission of ESA scheduled for 2015 (out of the ecliptic views of the polar regions of the Sun).

\end{document}